\documentclass[]{article}

 \usepackage[]{graphicx}
 \usepackage{epsfig}
 \usepackage{amsmath, amstext, amsfonts}
 \usepackage{amssymb}
 \usepackage{multicol}
 \usepackage[latin1]{inputenc}

\usepackage{anysize} 
\marginsize{2cm}{2.5cm}{1.5cm}{2cm}

 \title{Detection of mixed-culture growth in the total biomass data by wavelet transforms}
\author{H.C. Rosu$^1$, J.S. Murgu\'{\i}a$^2$, V. Ibarra-Junquera$^3$\\
$^1$ IPICyT (San Luis Potos\'{\i}, Mx)\\ $^2$ UASLP (San Luis Potos\'{\i}, Mx)\\ $^3$ UCOL (Colima, Mx)
}
 \date{J. Appl. Res. Tech. 8(2), 240-248 (2010)}

\begin{document}

 \maketitle
\begin{abstract}
  \footnotesize
   \noindent
   We have shown elsewhere that the presence of mixed-culture growth of microbial species in fermentation processes can be detected with high accuracy by employing the wavelet transform. This is achieved because the crosses in the different growth processes contributing to the total biomass signal appear as singularities that are very well evidenced through their singularity cones in the wavelet transform. However, we used very simple two-species cases. In this work, we extend the wavelet method to a more complicated illustrative fermentation case of three microbial species for which we employ several wavelets of different number of vanishing moments in order to eliminate possible numerical artifacts. Working in this way allows to filter in a more precise way the numerical values of the Hölder exponents.
Therefore, we were able to determine the characteristic Hölder exponents for the corresponding crossing singularities of the microbial growth processes and their stability logarithmic scale ranges up to the first decimal in the value of the characteristic exponents. Since calibrating the mixed microbial growth by means of their Hölder exponents could have potential industrial applications, the dependence of the Hölder exponents on the kinetic and physical parameters of the growth models remains as a future experimental task. \\
\\
\noindent Key-words: mixed-culture growth, total biomass, wavelets, Hölder exponents.\\
\noindent Source file: jart2010.tex\\
\\
\begin{center} {\bf Resumen}\end{center}
Hemos mostrado en un trabajo anterior que la presencia de crecimiento mixto de poblaciones en procesos de fermentación puede ser inferida por medio del análisis ondeleta, el cual detecta con mucha precisión cruces en la señal de biomasa total. Esto fue realizado debido a que los cruces en los diferentes procesos de crecimiento que contribuyen en la señal de biomasa total surjan como singularidades muy bien delineadas en sus conos de singularidad de la transformada ondeleta. Sin embargo, usamos casos muy simples de dos especies. En este trabajo extendemos el estudio de tipo ondeleta a un caso ilustrativo de fermentación más complicado de tres poblaciones microbianas, en el cual se usaron varios tipos de ondeletas de diferente número de momentos de desvanecimiento para evitar los posibles artefactos numéricos. Trabajando de esta forma nos permite filtrar de manera más precisa los valores numéricos de los exponentes de Hölder.
Por consiguiente, pudimos determinar los exponentes característicos de Hölder para las singularidades de cruce correspondientes a los procesos de crecimiento microbiano así como de sus rangos estables del logaritmo de la escala hasta el primer decimal en el valor de los exponentes característicos. Debido a que la calibración del crecimiento mixto microbiano por medio de los exponentes de Hölder podría tener aplicaciones industriales potenciales, la dependencia de los exponentes de Hölder en función de los parámetros cinéticos de los modelos de crecimiento queda como una futura tarea experimental.\\
\\
\noindent Palabras claves: crecimiento de cultivo mixto, biomasa total, ondeletas, exponentes de Hölder.\\
\noindent Archivo fuente: jart2010.tex
 \end{abstract}


\section{Introduction}
The growth of microbial species in media containing two or more substrates limiting the growth is of considerable biotechnological and bioengineering interest \cite{India}. The mixed
growth of microorganisms occurs in many industrial processes. An important class of such processes is that of the traditional fermentation food and drinks in which either ambient microorganisms or inoculums with selected microorganisms are used. The presence of different microbial species and substrates is a dominant factor for the quality and quantity of the final product. Biomass measurements are among the most important measurements in any cell culture process and detecting the presence or absence of the mixed culture growing processes is a demanding issue. In general terms, the biomass measurements can by divided in continuous and on-line, and discrete and off-line. The continuous biomass acquisition signals and reliable on-line methods yield valuable knowledge on the status of the process and can facilitate the process monitoring and control. Several methods have been developed for the continuous measurement of biomass. Traditional optical methods such as absorbance measurements are probably the easiest ways and are the most commonly used. Additionally, capacitance has been also used to estimate the biomass concentration in a continuous and online way, see for instance the web pages of the Fogale and Applikon companies: http://www.fogale.fr/biotech/pages/yeasts-2.php and http://www.applikonbio.com/applikonbio/c8-1.htm, respectively.
Our main goal is to show that it is possible to infer the mixed culture microbial growth based only on the knowledge of total biomass data, without employing complicated techniques. The alternative procedure that we emphasized recently and it is here shortly presented is to apply the wavelet method to the fermentation data with the purpose to detect singularities in the growth curves (signals). In this case, one can consider the mixed culture growth curves as more or less regular signals containing crosses due to the difference in growth of the different species. In the wavelet literature, one can find fundamental papers in which it is shown that the wavelet techniques provide an efficient tool for detecting singularities, see for example Mallat and Hwang \cite{MH}.

\section{Microbial kinetics}

The fermentation processes on which we focus here evolve in stirred batch tank reactors. In such a case, the biomass converts the substrate in additional biomass and products. The dynamical model of the bioreactor for biomass production is given by the following ordinary differential equations:
\begin{eqnarray}
\frac{ \mathrm{d} x_{1,i}}{ \mathrm{d} t}  &=& \  x_{1,i} \  \mu _i\left( x_{2,i} \right) \nonumber\\
\frac{ \mathrm{d} x_{2,i}}{ \mathrm{d} t}  &=& \
-\frac{x_{1,i}}{Y_i}\  \mu _i \left(x_{2,i} \right) \nonumber
\end{eqnarray}
$x_{1,i}$ are the biomass concentrations, $x_{2,i}$ the concentrations of the substrate, $Y_i$ is the yield factor defined as the ratio between the amount of produced biomass per unit of time and the amount of consumed substrate per unit of time for each species, $\mu_{i}(x_{2,i})$  are the growth rates of each species and the subindex  $i$  represents the number of species and substrates involved in the fermentation process. The growth rates relate the changes in the biomass concentrations with the (negative) changes of the substrate concentrations per unit of time. We will use the formulas of Monod and Haldane, respectively:\\

1.-	For the substrate saturation model (the Monod equation):
\begin{eqnarray}
\mu _i\left( x_{2,i}\right)\ = \ \frac{\mu_{max_i}\
x_{2,i}}{K_{{1}_i} + x_{2,i}}\nonumber
\end{eqnarray}
where $K_{1_i}$ is the saturation (or Monod) constant.\\

2.-	For the substrate inhibition model (the Haldane equation):
\begin{eqnarray}
\mu _i\left( x_{2,i}\right)\ = \ \frac{\mu_{max_i}\
x_{2,i}}{K_{{1}_i} + x_{2,i} + K_{{2}_i}x_{2,i}^2}\nonumber
\end{eqnarray}
where in addition to the saturation constant there is an inhibition constant $K_{2_i}$.\\

In both models, $\mu_{max_i}$  is the maximal specific growth rate. For simplicity, this parameter has been taken as unity in the illustrative case presented in the following. The value of $K_{1_i}$ expresses the affinity of the biomass for the substrate. The Monod growth kinetics can be considered as a special case of the substrate inhibition kinetics for $K_{2_i}=0$, i.e., when the inhibition terms vanish. The total biomass signal (TBS) is naturally given by
\begin{eqnarray}
y(t) = \ \sum_{i=1}^{m} x_{1,i}
\end{eqnarray}
where $m$ is the total number of microbial species inside the
bioreactor. In our previous work \cite{previous}, we considered the
simple case of two species and two substrates. In Section 3, we will present a more complicated case, namely the case of three species, i.e., $m=3$, with identical initial conditions for all species and substrates.

\section{Wavelet analysis}

In the context of signal theory the total biomass signal $y(t)$ is projected onto a wavelet basis through the correlation of the signal with an integral kernel which is a scaled and translated version of an analyzing wavelet
\begin{equation}\label{3-1}
T_\psi[y](a,b)=\frac{1}{\sqrt{a}}\int _{-\infty}^{\infty}y(t)\psi\left(\frac{t-b}{a}\right)dt~,
\end{equation}
where $\psi$ is the analyzing wavelet. The values of Eq.~(\ref{3-1}) are called the wavelet coefficients of the signal y(t). In general, a signal can be disentangled into three parts: (a) a mostly smooth continuous structure that can be represented by low-order (piecewise) polynomials, (b) a discontinuous part intrinsic to the particular nature of the signal, and (c) a noise part which can also be quite discontinuous.
In general, for any function $f(t)$  we can define its moments as $M_k=\int _{-\infty}^{\infty}t^kf(t)dt$, for $k$ a nonnegative integer. In the case of wavelets, an analyzing wavelet $\psi$ has $p$ vanishing moments if and only if it satisfies
\begin{equation}\label{3-2}
\int _{-\infty}^{\infty}t^k\psi(t)dt=0~, \qquad {\rm for}\, {\rm all}\,\, 0\leq k < p~.
\end{equation}
This property means that the analyzing wavelet $\psi$ is orthogonal to any polynomial having degree up to $p-1$. The closer a signal is to being orthogonal to the analyzing wavelet, the smaller the resulting wavelet coefficient. Thus, if the analyzed signal can be approximated by a low-order polynomial over the support of the wavelet, then the wavelet is nearly orthogonal to the signal and the resulting wavelet coefficients will be very small (resulting from the approximation error). The smaller the wavelet support, the greater the precision in detecting discontinuities and the shorter the processing time. Moreover, the support of a wavelet is directly proportional to its number of vanishing moments. We will display results presenting the same total biomass signal analyzed with wavelets of the Gaussian family for different vanishing moments, from two to four. In addition, as in our previous work \cite{previous}, we use the wavelet transform modulus maxima procedure introduced by Mallat and Hwang \cite{MH} that provides a precise location of the singularity and a clear cut measure of its Hölder exponent (HE) from the scaling of the wavelet transforms along the so-called modulus maxima lines on which the transforms reach local maxima with respect to the position coordinate. For more details, the reader is directed to the works \cite{MH} and \cite{previous}.

\section{ A case study: Two Haldane and one Monod species}

Depending on the saturation and inhibition parameters and the types of fermentative growth, the crosses in the case of three growing species can occur at well-separated moments. For example, the total biomass signal may have two separated cones of influence in its wavelet transform. Such a case for two Haldane populations and one of Monod type is displayed in the figures of this work for different analyzing wavelets. Processing the same signal with several analyzing wavelets is recommendable when a better identification of the HEs is sought and helps to get rid of all sorts of artifacts. We used the second, third, and fourth derivatives of the Gaussian function as analyzing wavelets. It is well known that these Gaussian derivative wavelets have the same number of vanishing moments as the order of the derivative. In each case the range of scales with the same HE up to the first decimal digit in all three continuous wavelet transforms have been found. The same type of analysis, namely using wavelets of different vanishing moments to sense a given signal, have been used by Arneodo {\em et al} \cite{arne95} in a research of the fractal properties of DNA genome sequences. The argument for working in this way is that the common features showing up in all wavelet transforms belong with much higher confidence to the signal itself and it is not a numerical artifact caused by the integral transform. Following this idea, in the case of three microbial populations presented here, we can conclude that the HE is 1.6 for the first crossing singularity and 1.3 for the second one. This is so because these values are preserved in sufficiently long scale ranges although not for the entire scale interval that we considered.

\begin{itemize}

\item \underline{The first crossing singularity}\\
The results concerning the first encountered singularity are displayed in the figures 1, 2, and 3. From the plots ($c$) in each of the figures, we notice that we are able to get a stable first decimal digit of the HE if the low scales in the calculation of the wavelet transform are eliminated roughly proportional with the order of the Gaussian derivative employed.

\begin{figure}[!thp]
\begin{center}
\includegraphics[height=76mm]{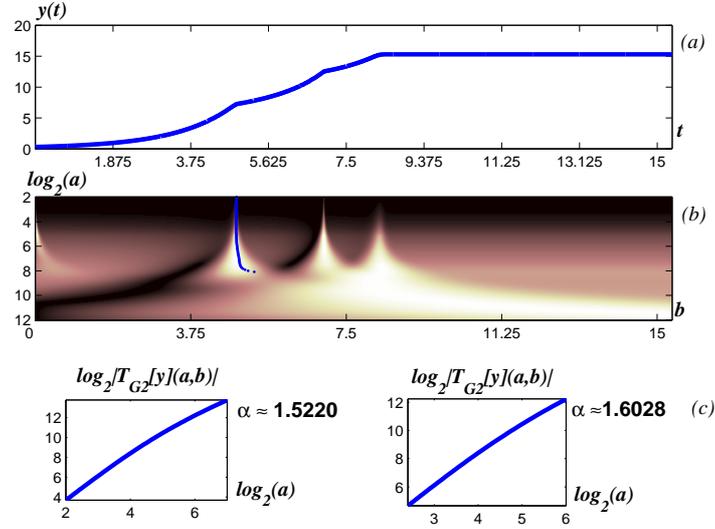}
\caption{ ($a$) Total biomass signal (TBS) for three microbial species with different growth rates: two are of Haldane type with  $K_{1_1}= 0.02$ g/l and  $K_{2_1}=0.04$ g/l, and  $K_{1_2}=0.03$ g/l and  $K_{2_2}=0.9$ g/l, respectively and the third one is of Monod type with $K_1 = 0.09 g/l$. ($b$) The singularity analysis has been performed with the second Gaussian derivative wavelet. ($c$) The HE $\alpha$ is shown in two different scale ranges.}
\label{fig:1G2}
\end{center}
\end{figure}

\begin{figure}[!thp]
\begin{center}
\includegraphics[height=76mm]{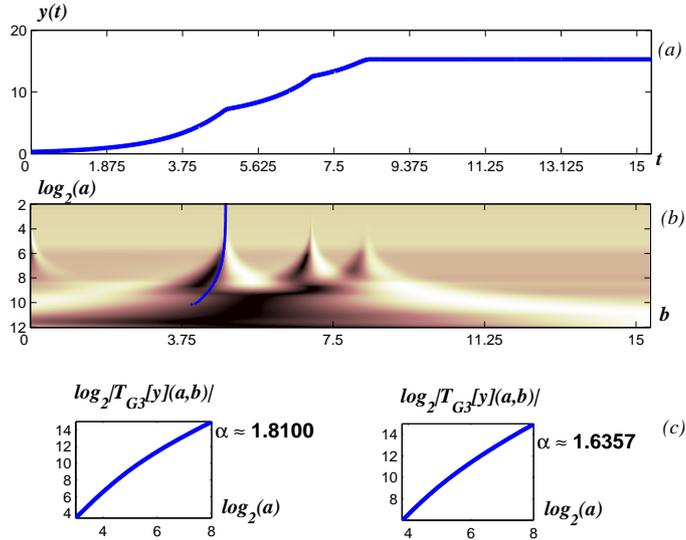}
\caption{The same case as in Fig. 1 but with the singularity analysis performed with the third Gaussian derivative wavelet.}
\label{fig:1G3}
\end{center}
\end{figure}

\begin{figure}[!thp]
\begin{center}
\includegraphics[height=76mm]{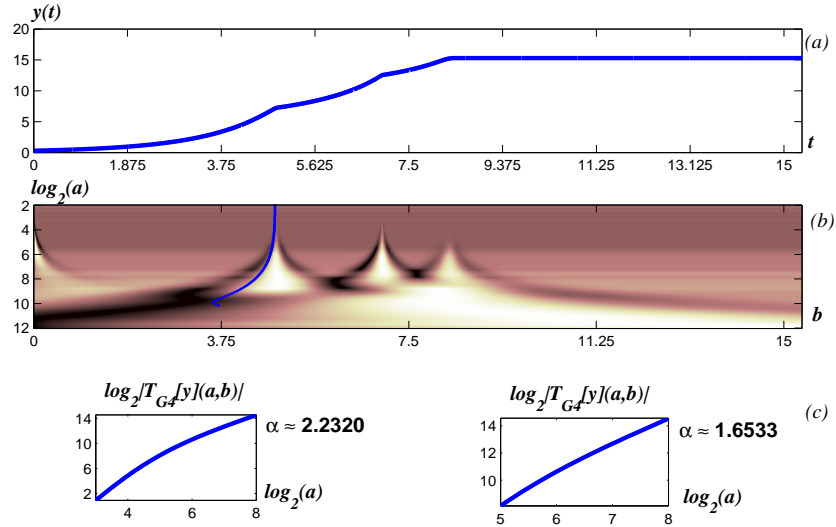}
\caption{ The same case as in previous figures but with the singularity analysis performed with the fourth Gaussian derivative wavelet.}
\label{fig:1G4}
\end{center}
\end{figure}

\item \underline{The second crossing singularity}\\
The results concerning the second singularity are displayed in table~1 (we do not include the plots for this singularity here). In this case, we could not assess a similar trend for the low scales. However, we identified the scale ranges on which we got a stable first decimal digit of the exponent for the three Gaussian derivatives, as one can see from the right plots (c).

\item \underline{The plateau singularity}\\
The plateau singularity is less interesting because it indicates only the end of the fermentation process. The only useful fact is its detection and we noticed that it is indeed detected with any of the employed wavelets.

We collect the results of our illustrative example in table 1. One can notice that we can determine a range of scales where the HE is the same up to the first decimal digit independently of the number of vanishing moments of the used wavelet. In the case of the first crossing, HE is constant only over two log scales (5-6), whereas for the second crossing it stays constant over three log scales (3-5).\\

\begin{table}[h]
\begin{center}
\begin{tabular}
[b]{|l|l|l|l|}\hline
Type of wavelet & 1st crossing & 2nd crossing & Plateau \\\hline
Gaussian derivatives & $R_{1f}\quad $ \vline\vline $\quad R_{2f}$ & $R_{1s}\quad $ \vline \vline $\quad R_{2s}$ & Detected ?\\\hline
$G2$ & $1.5220\,\,$\vline $\,\,1.6028$ & $1.1600\,\,$\vline $\,\,1.3200$ & $\quad$Yes\\\hline
$G3$ & $1.8100\,\,$\vline $\,\,1.6357$ & $1.2440\,\,$\vline $\,\,1.3730$ & $\quad$Yes\\\hline
$G4$ & $2.2320\,\,$\vline $\,\,1.6533$ & $1.2429\,\,$\vline $\,\,1.3514$ & $\quad$Yes\\\hline
\end{tabular}
\caption{The Hölder exponents obtained in two different ranges of scales $R_1$ and $R_2$ for the first (subindex $f$) and the second (subindex $s$) crossing singularities, respectively,  with the wavelet functions corresponding to the second (G2), third (G3) and fourth (G4) derivatives of the Gaussian function.}
\end{center}
\label{tab:1}
\end{table}

As a general remark, notice that we did not use the first Gaussian derivative as an analyzing wavelet because in this case the HEs are very close to unity which means that the signal is continuous at the singularity detected with higher-order derivative Gaussian wavelets; this fact tells us that the continuous wavelet transform with this wavelet cannot detect the singularities in the signal independently of the range of scales that one takes into account, see also Mallat and Hwang \cite{MH}.

\end{itemize}

\section{Conclusions}

The presence of mixed culture growth in fermentation processes can be inferred starting from the data of total biomass processed with the wavelet transform acting as detector of the crossing growth singularities.

\begin{itemize}

\item 	The most relevant quantity provided by the wavelet analysis is the HE of the singularities that we determined here through the illustrative example of three microbial species contributing to the fermentation.

\item	In order to have a real technological application of the wavelet techniques in this area, it is important to perform calibrations of the mixed-culture growth in terms of their HEs.

\item	Elimination of possible numerical artifacts can be achieved by employing wavelets of different vanishing moments for analyzing the same set of data.  For this goal, we used the Gaussian derivative wavelets, which have the same number of vanishing moments as the order of the derivative. It is known from the wavelet theory that working with analyzing wavelets with $p$ vanishing moments provide information on the regularity of the $(p-1)$th derivative of the analyzed signal. Since the majority of the numerical values of the HEs that we obtained are between one and two, it is clear that the fermentation signal under investigation has singularities in the first derivative. At the same time, this explains why we did not employed the first Gaussian derivative wavelet since it has less vanishing moments than the numerical value of the HEs.

\item	It would be very interesting to use other recent techniques to measure the fractality of the total biomass signals. We mention the DFA (Detrended Fluctuation Analysis) \cite{Peng} and the MF-DFA (Multifractal-DFA) \cite{kant}. These methods are based on determining the scaling properties of the fluctuating moment of order $q$ ($q=2$ holds for the case of DFA), as a function of the size of the samples and when the data contain noises and trends that are unknown from the point of view of their origin and their specific forms.
In addition, one can consider biological growths whose parameters depend on specific physical parameters, such as the temperature, which is common for long-term bacterial activities \cite{price}. Once the signal depends on the temperature, the HE depends as well. It would be highly interesting to examine the temperature dependence of the HE as implied by the temperature dependence of the total biomass signal.

\end{itemize}


\begin{thebibliography}{99}
\bibitem {India} Joshi V. K. \& Pandey A., eds.,
Biotechnology: Food Fermentation (Microbiology, Biochemistry and Technology), Vol. 2, New Delhi, Educational Publisher \& Distributors, 524-1372 (1999).

\bibitem{MH} Mallat S. \& Hwang W.L.,
{\em Singularity detection and processing with wavelets},
IEEE Trans. Inf. Theor. {\bf 38}, 617-643 (1992).

\bibitem{previous} Ibarra-Junquerra V., Escalante-Minakata P., Murguía J. S. \& Rosu H. C.,
{\em Inferring mixed culture growth from total biomass data in a wavelet approach},
Physica A {\bf 370}, 777-792 (2006).

\bibitem{arne95}
            Arn\'eodo A., Bacry E., Graves P.V. \& Muzy J.F.,
{\em Characterizing long-range correlations in DNA sequences from wavelet analysis},
            Phys. Rev. Lett. {\bf 74}, 3293-3296 (1995).

\bibitem{Peng} Peng C.-K., Buldyrev S. V., Havlin S., Simons M., Stanley H. E. \& Goldberger A. L.,
{\em Mosaic organization of DNA nucleotides},
Phys. Rev. E {\bf 49}, 1685-1689 (1994).

\bibitem{kant} Kantelhardt J. W., Zschiegner S. A., Koscielny-Bunde E., Havlin S., Bunde A. \& Stanley H. E.,
{\em Multifractal detrended fluctuation analysis of nonstationary time series},
Physica A {\bf 316}, 87-114 (2002).

\bibitem{price} Price P. B. \& Sowers T.,
{\em Temperature dependence of metabolic rates for microbial growth, maintenance, and survival},
Proc. Nat. Acad. Sci. USA {\bf 101}, 4631-4636 (2004).

\end{thebibliography}
\end{document}